\documentclass[preprint,showpacs,preprintnumbers,amsmath,amssymb]{revtex4-1}
\usepackage{graphicx}
\usepackage[draft,inline]{fixme}
\usepackage{amsmath} 
\usepackage{amsfonts} 
\usepackage{amssymb} 

\begin{document}
\draft
\title{Stamp transferred suspended graphene mechanical resonators for radio-frequency electrical readout}

\author{Xuefeng Song$^\dag$}
\author{Mika Oksanen$^\dag$}
\author{Mika A. Sillanp\"a\"a$^\dag$}
\author{H. G. Craighead$^\ddag$}
\author{J. M. Parpia$^\ddag$}
\author{Pertti J. Hakonen$^{\dag,}$\footnote{Corresponding author. E-mail: pjh@boojum.hut.fi.}}
\affiliation{$^\dag$ Low Temperature Laboratory, School of Science, Aalto University, P.O. Box 15100, FI-00076 Aalto, Finland}
\affiliation{$^\ddag$ Center for Materials Research, Cornell University, Ithaca, New York 14853, USA}

\begin{abstract}
We present a simple micromanipulation technique to transfer suspended graphene flakes onto any substrate and to assemble them with small localized gates into mechanical resonators.  The mechanical motion of the graphene is detected using an electrical, radio-frequency (RF) reflection readout scheme where the time-varying graphene capacitor reflects a RF carrier at $f=5-6$ GHz producing modulation sidebands at $f \pm f_{m}$.  A mechanical resonance frequency up to $f_m=178$ MHz is demonstrated.  We find both hardening/softening Duffing effects on different samples, and obtain a critical amplitude of $\sim40$ pm for the onset of nonlinearity in graphene mechanical resonators. Measurements of the quality factor of the mechanical resonance as a function of DC bias voltage $V_{DC}$ indicate that dissipation due to motion-induced displacement currents in graphene electrode is important at high frequencies and large $V_{DC}$.

\end{abstract}


\maketitle
Graphene is a perfect two dimensional crystal with high Young's modulus $E \sim$ 1 TPa \cite{lee2008measurement} and exteremly low mass, which makes it ideal for high-frequency, high-\emph{Q} nano-electromechanical systems (NEMS). It is a promising material for extremely low mass sensors \cite{Jensen2008NatureNanotechnology,Naik2009NatureNanotechnology,Ekinci2005} and for mechanical resonators at the quantum limit \cite{Schwab2005PhysicsToday,Aspelmeyer2008,ClelandMartinis}.  An experimental demonstration of a monolayer graphene resonator using optical methods \cite{bunch2007electromechanical} was achieved soon after the discovery of this material \cite{Novoselov2004}. Compared with optical methods \cite{bunch2007electromechanical,Barton2011NanoLetters}, electrical detection schemes \cite{Chen2009NatureNanotechnology,Eichler2011NatureNanotechnology} are more compatible with microelectronic applications and, moreover, they facilitate easier studies of fundamental phenomena at low temperatures, where higher \emph{Q} values and better sensitivity are obtainable.

Several methods for measuring and fabricating high frequency NEMS have been developed \cite{Schwab2005PhysicsToday,LaHaye2004Science,Clerk2010RMP,Sulkko2010,Huang2003,Sazonova2004Nature,Garcia-Sanchez2007,Li2008,Regal2008,Sillanpaa2009}. They all share the challenges brought by downsizing towards the submicron-scale, since the vibration amplitude under constant drive force of the mechanical bar-type resonator diminishes proportional to length squared \cite{Clelandbook,Ekinci2005}. The optimization of a NEMS-resonator geometry is dependent essentially on the read-out scheme to be employed. When the use of capacitive techniques with \emph{LC} matching circuits \cite{Sillanpaa2009,Sulkko2010} is envisaged, in most cases, decreasing the parasitic capacitance is crucial for increasing sensitivity.  Unfortunately, using common preparation methods for suspended graphene samples, such as undercut etching of a sacrificial layer (\emph{e.g.} SiO$_2$) \cite{Bolotin2008SolidStateComm} or random exfoliation over predefined trenches \cite{garcia2008}, it is hard to make small-capacitance, localized gates for graphene mechanical resonators. Generally speaking, the undercut etching leads to high parasitic capacitance\cite{Singh2010Nanotechnology}, while random exfoliation suffers from low success rate.  There have been some pioneering attempts on transferring carbon nanotubes \cite{Kang2007NatureNanotechnology} and graphene pieces \cite{Zomer2011Arxiv,Reina2008NanoLetters, Dekker2010NanoLetters,Hone2010NatureNanotech,Bie2011AdvancedMaterials,Castellanos2011Small} with polymer films, but realizing suspended structures after transfer remains difficult. In this work, we have developed the polymer transfer method into a micron-scale, e-beam patterned stamp technique, which allows us to move individual suspended graphene flakes and assemble them with small localized gates into electrically-controlled mechanical resonators.  We demonstrate, for the first time, the dispersive readout scheme for graphene mechanical resonators.

\begin{figure}[h]
  \includegraphics[width=8.5cm]{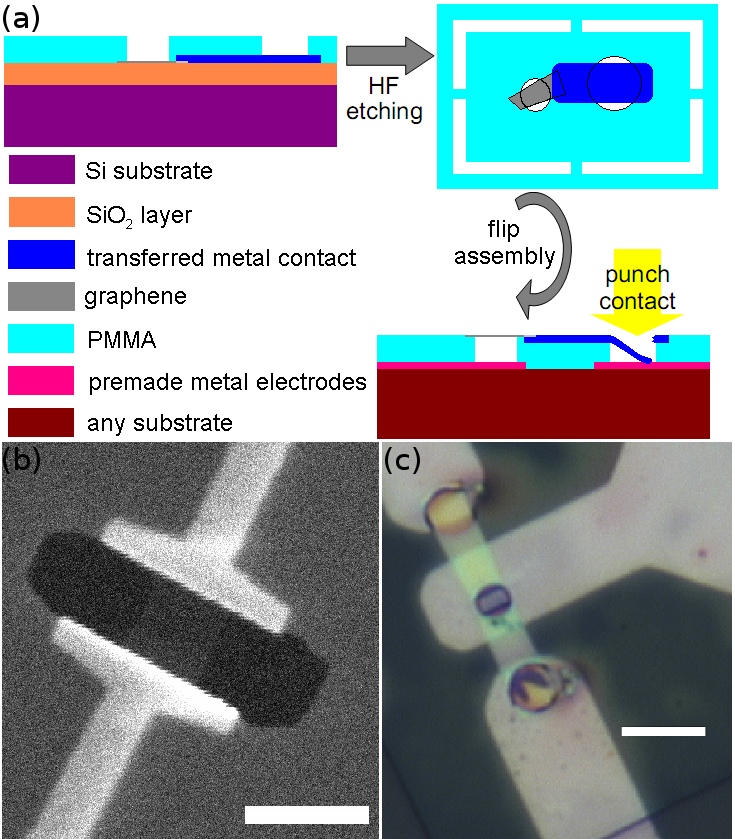}
  \caption{Fabrication of suspended graphene mechanical resonators. (a) Schematic drawing of our suspended stamp transfer technique. (b) SEM image of a suspended graphene in a PMMA stamp; the scalebar equals $1\ \mu$m. (c) Optical image of a graphene mechanical resonator on a sapphire substrate with localized gate electrode; the scalebar corresponds to $5\ \mu$m. } \label{StampTech}
\end{figure}
The basic steps of our suspended stamp technique are shown in Fig. \ref{StampTech}(a).  First, high quality tape-exfoliated monolayer graphene pieces are located on a silicon substrate covered with 275 nm thick SiO$_2$ (confirmed with Raman spectroscopy).  50 nm gold electrodes are deposited on the graphene pieces by a normal e-beam lithography (EBL) and lift-off procedure.  Then PMMA is spun on the chip again and a 2nd EBL step is applied to define patterns such as stamps and holes in it.  Next, we peel off the whole PMMA membrane ($\sim 4\textrm{mm}\times 4\textrm{mm}$) from the initial substrate by etching away the SiO$_2$ in 1\% HF solution, rinse it with DI water, and fish the whole membrane out of water using a ring-shaped frame.  The graphene pieces and gold electrodes are both embedded in the PMMA stamps, which are linked with the whole PMMA membrane by some weak joints on the edges, as shown in Fig. \ref{StampTech}(a). In fact, the graphene pieces are clamped by the gold electrodes and suspended across the holes in the PMMA stamps, as shown in Fig. \ref{StampTech}(b). We can locate any individual stamp under an optical microscope. Then using a fine-tipped glass needle controlled by a micromanipulator \cite{Song2009Nanotechnology}, we can break the joints to pick up a specific stamp, transfer and lay it onto any desired target spot with a precision of $\sim1\ \mu$m. Although the suspended monolayer graphene pieces are invisible under the optical microscope, the position can be well judged by the visible electrodes and holes in the stamp.  During the transfer, we invert the stamp thus bringing the graphene and electrodes from the bottom to the top of the stamp, so that they are supported by the PMMA and will be kept suspended over the gate after assembly. The typical gap distance between the gate and the suspended graphene is 0.1-0.5 $\mu$m, which is controlled by the thickness of the PMMA. Finally, to make electrical contact between the hanging gold electrodes and the premade electrodes of the target circuit, we press the hanging electrodes through windows in the PMMA down to punch contact pads using the same glass needle. The punch method results in good ohmic gold-gold contacts in our experiments. We make the on-chip target electrodes via normal EBL using Ti/Au deposition on a sapphire substrate. The suspended graphene is grounded via punch contacts and the localized gate below it is bound to the signal line. A typical optical image of our sample is shown in Fig. \ref{StampTech}(c).  Since the graphene pieces are suspended during the entire store and transfer process, the success rate depends on the size of the suspended part.  At present, our success rate for samples below 2 $\mu$m is about 50\%.

\begin{figure}
\includegraphics[width=\columnwidth]{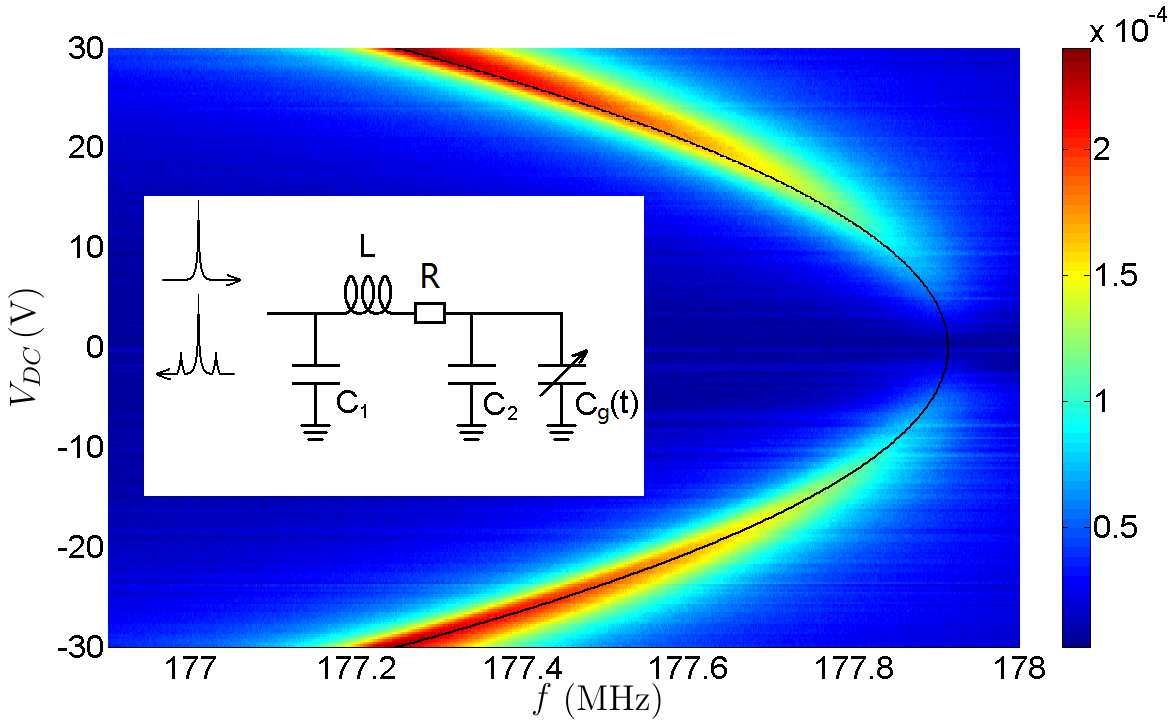}%
\caption{Mechanical resonance frequency $f_m$ for sample \#1 measured at 4.2K. The amplitude of the sideband $V_{\pm}$ is given by the color scale in Volts as a function of driving frequency $f$ and DC gate voltage $V_g$. The solid line is the fitted $f_m(V_{DC})$ parabolic curve from equation (4). Inset: the sideband reflection measurement scheme using a $\pi$ matching circuit.} \label{Sample178MHzAndCircuit}
\end{figure}

Cavity-based capacitive readout methods have proven very efficient in studying micromechanical resonators, but their application becomes increasingly difficult when the operating frequency is increased towards 1 GHz \cite{Sillanpaa2009}. Compared with mixing techniques, the loss in sensitivity due to larger measurement frequency is compensated by the possibility of working without coherent excitation for driving the mechanical oscillations.  The main problem in high frequency readout is stray capacitance that easily masks the variation in capacitance $\delta C \ll 1$ fF induced by mechanical vibrations. Parasitic capacitance can be eliminated to a high degree by tuning the circuitry by an on-chip inductance \cite{Sulkko2010}, but even then the sensitivity may remain moderate because of the stray capacitance of the inductor itself. A detailed analysis of the ideas involved in the RF capacitive readout method can be found in a recent review \cite{Clerk2010RMP}.  In this work, we have employed a $\pi$-matching network where the Al bond wire forms the inductive element, in a manner shown in the inset of Fig. \ref{Sample178MHzAndCircuit}. The $\pi$-matching circuit enhances the sensitivity of the measurement, but the actual result is very sensitive to the bond wire length and the optimization must be done by trial and error.

For a clamped-clamped graphene mechanical resonator with length \emph{L}, the time varying average gap between the gate and the graphene can be written as $D(t)=D_{eq}+\cos(\omega_m t) \frac{1}{L} \int_0^L A(y)dy$ when vibrating at a mechanical resonance frequency $f_m = \omega_m /2\pi$.  Here $D_{eq}$ is the average gap at the equilibrium position, and $A(y)$ denotes the amplitude of the flexural mode.  The oscillation of the gap width gives rise to a time-varying capacitance $C_g(t) = C_{eq} + \delta C\cos(\omega_m t)$ where $\delta C = \frac{dC}{dD} \frac{1}{L} \int_0^L A(y)dy$. According to lumped-element modeling of the $\pi$-matched \emph{LC} resonator, the time varying $C_g(t)$ leads to an observable frequency modulation of the tank circuit down to $\delta C \sim 0.1$ aF.
In our RF reflection measurement, the voltage applied on the input of the $\pi$ circuit is $$V(t) = V_{DC} + V_{AC} \cos(\omega_m t) + V_{LC} \cos(\omega_{LC} t). \eqno{(1)}$$  Here $ V_{DC}$ is the constant voltage bias on the graphene capacitor, $V_{AC} $ actuates the membrane at $\omega_m$, and $V_{LC}$ marks the probe RF signal at $\omega_{LC}$. The impedance modulation by the capacitance leads to side peaks $V_{\pm}$ in the reflected power at $\omega_{LC} \pm \omega_{m}$. In our case, $C_1=1.5$ pF, $C_2\approx50$ fF, $L\approx15$  nH, and the fitting of a measured tank resonance at $f_{LC}=5.89$ GHz yields $Q\approx80$ corresponding an overall resistance of $R\approx6$ $ \Omega$.  Using circuit analysis with these parameters and the parallel plate approximation for the graphene capacitor, we obtain $$\frac{V_{\pm}}{V_{LC}}\approx (4.0\times 10^3 \textrm{m}^{-1} \frac{S}{D_{eq}}) \frac{\int_0^L A(y)dy}{D_{eq}}=(4.0\times 10^3 \textrm{m}^{-1} \frac{S}{D_{eq}}) \frac{\delta D}{D_{eq}} \eqno{(2)}$$ for the relation between the graphene vibration amplitude and the sideband voltage. Here $\delta D=\frac{1}{L} \int_0^L A(y)dy$ is the average deflection representing the mechanical vibration amplitude and we have assumed that $\frac{dC}{dD}$ is independent of position along the resonator.  In our experiments, $\delta D \ll D_{eq}$, $\delta C \ll C_{eq}$.

The electrical force acting on the resonator is $$F(t) = \frac{1}{2} \frac{\partial C}{\partial D }|_{D=D_{eq}} V^2(t). \eqno{(3)}$$  When $|V_{DC}|$ is small, $D_{eq}$ and the tension in the graphene do not change much with $V_{DC}$. Without tension the effective spring constant becomes $K_{eff}(V_{DC})=K_0-\frac{1}{2}\frac{\partial^2 C}{\partial D^2}|_{D=D_{eq}} V_{DC}^2$, where $K_0$ is the intrinsic spring constant of the graphene resonator.  Consequently, there is a parabolic bias dependence given by $$f_m(V_{DC})=f_{m0}(1-\gamma V_{DC}^2), \eqno{(4)}$$ where $\gamma=\frac{1}{4K_0}\frac{\partial^2 C}{\partial D^2}|_{D=D_{eq}}$ is a constant governed by the geometry of the resonator as well as the non-idealities of the graphene sheet, \emph{e.g.} rippling and edges.

The measurements were carried out in a high vacuum chamber dipped into liquid Helium (4.2 K). The actuator signal was taken from the internal reference of SR844 RF lock-in (up to 200 MHz). The RF carrier signal at $\omega_{LC}$ was injected through a circulator located on top of the cryostat to the sample, and the reflected sidebands were amplified using $Miteq$ low noise amplifiers (band 4-8 GHz). After down-mixing to the actuator frequency, the sideband amplitude was recorded by the RF lock-in using a time constant $\tau_{lock-in}=30$ ms.  Fig. \ref{Sample178MHzAndCircuit} displays the measured sideband voltage as a function of the actuator frequency $f$ and the DC bias voltage $V_{DC}$, with the highest mechanical resonance frequency around 178 MHz obtained on sample \#1 with length $L=0.7\ \mu$m and width $W=1\ \mu$m. The concave parabolic $f_m$ dependence on $V_{DC}$ fits well Eq. (4).

\begin{figure}
\includegraphics[width=\columnwidth]{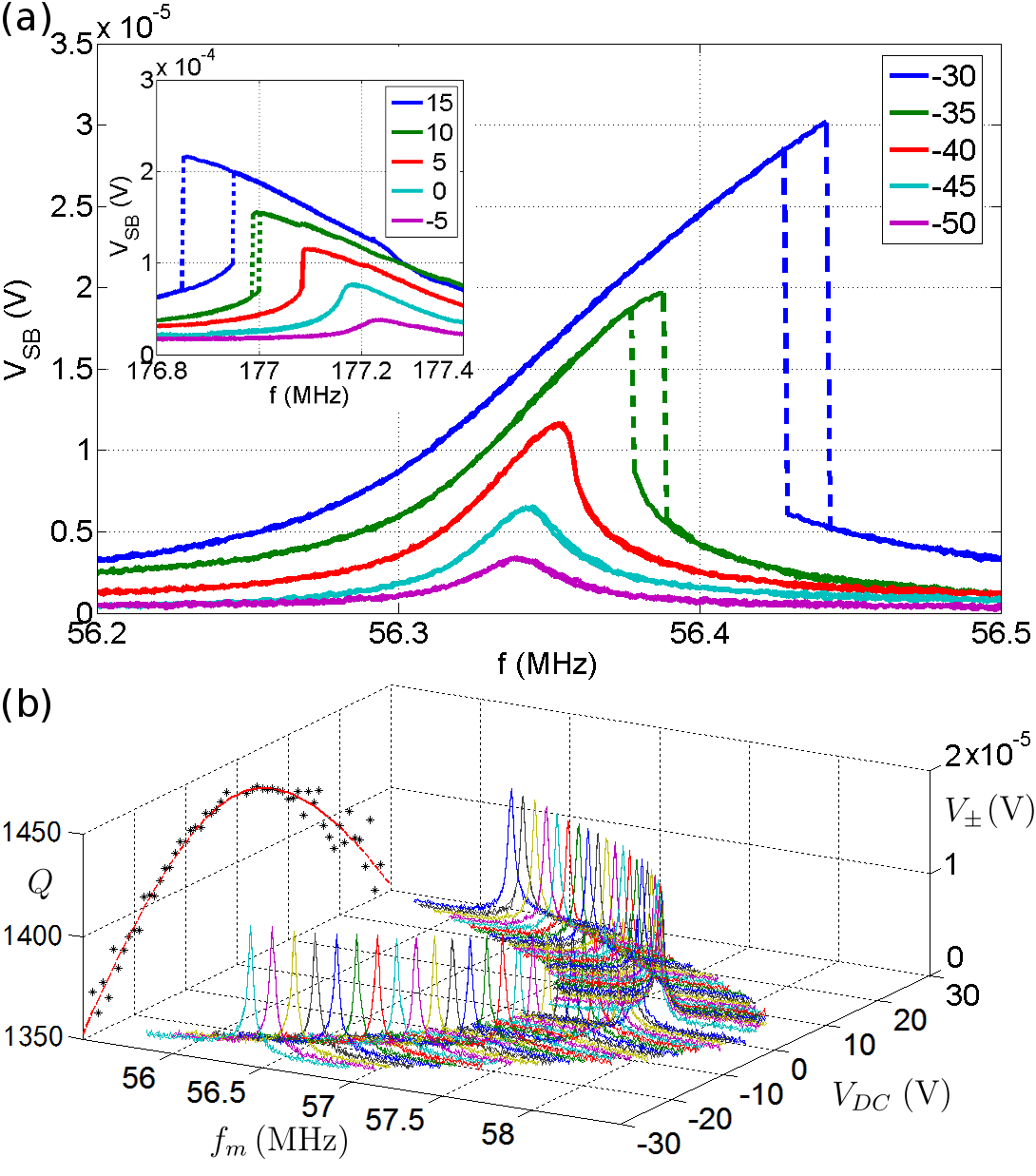}%
\caption{(a) Hardening Duffing behavior sample \#2 around 56.3 MHz (with $V_{DC} = -30$ V and drive generator power stepped from -50 dBm to -30 dBm). Inset: softening Duffing effect addressed on sample \#1 around 177 MHz (with $V_g = 30$ V and generator power increasing from -5 dBm to 15 dBm). (b) A series of resonance curves of a graphene mechanical resonator under constant driving and the corresponding $Q(V_{DC})$ dependence which fits Eq. (5), the displacement current dissipation model (red dashed line). }\label{DuffingAndQ}
\end{figure}

From Eqs. (1) and (3), the driving force at $\omega_m$ is proportional to $ \frac{\partial C}{\partial D }|_{D=D_{eq}} \cdot V_{DC} \cdot V_{AC}$. In our experiments, nonlinear resonance behavior was found on all samples at large drives. Fig. \ref{DuffingAndQ}(a) shows a typical set of resonance curves displaying hardening Duffing behavior, when we change $V_{AC}$ at constant $V_{DC}$. Contrary to other samples, softening Duffing effect was observed on sample \#1, as shown in the inset of Fig. \ref{DuffingAndQ}(a). The hardening/softening behaviors correspond to the sign of the coefficient $\alpha_3$ of the restoring force $\alpha_1 u(t) + \alpha_3 u^3(t)$ in the Duffing equation, where $u(t)$ is the vibration displacement.  Due to the competition between elastic and capacitive mechanisms \cite{KozinskyAPL2006}, $\alpha_3$ can be either positive or negative, resulting in the observed hardening and softening Duffing behaviors, respectively.  The dependence of $\alpha_1$ on the capacitance derivatives and intrinsic elastic parameters is different from that of $\alpha_3$, and thus, there is no clear cut connection between $f_m(V_g)$ and hardening/softening Duffing behavior. However, the actual criterion for the cross-over between hardening and softening Duffing regimes is still unclear to us.  From Fig. \ref{DuffingAndQ}(a), with $L=1.5 \ \mu$m, $W=2\ \mu$m and $D_{eq}\approx 500\ \textrm{nm}$ for sample \#2, we can deduce the critical vibration amplitude $\delta D_{hyst}$ when the hysteresis emerges. For the red resonance curve in Fig. \ref{DuffingAndQ}(a), we have $V_{\pm}/V_{LC}\simeq 9.5 \times 10^{-7}$ which according to Eq. (2), yields for the critical average vibration amplitude $\delta D_{hyst}\approx 20\ \textrm{pm}$.  Using the approximations $A(y)=\frac{1-cos(2 \pi y/L)}{2}A_{max}$ and $\delta D=\frac{1}{L} \int_0^L A(y)dy$ \cite{KozinskyAPL2006} for the normalized basic mode shape, we obtain $A_{max}=2 \delta D$, where $A_{max}=max(A(y))$ is the antinode amplitude of the mechanical resonator with length $L$. So the measured critical amplitude for the real graphene resonator is about 40 pm, which agrees quite well with the  theoretical prediction of 50 pm \cite{Atalaya2008NanoLetters}. On the same sample, the height of the minimum detectable resonance peak is about 4\% of the hysteresis onset amplitude, which corresponds to 1.6 pm in the total vibration resolution. Taking into account the bandwidth of our measurement system $B=1/\tau_{lock-in} \approx 33$ Hz, we get for the sensitivity of our measurement $\sqrt{S_x}=1.6 \ \textrm{pm}/\sqrt{B}\approx 0.3\ \textrm{pm}/\sqrt{\textrm{Hz}}$, which is $\sim 10^2$ larger than the value of $\sim \textrm{fm}/\sqrt{\textrm{Hz}}$ achieved in the most sensitive detection schemes using SET or SSET \cite{Knobel2003Nature,LaHaye2004Science}.  For the fundamental mode of sample \#2, where $f_m\approx 57$ MHz and $Q\approx 1400$, the estimated RMS amplitude of its thermal motion at 4K is $A_{rms}^{th}=\sqrt{\frac{k_B T}{m_{eff} (2\pi f_m)^2}} \approx 4.4\ \textrm{pm}$.  Using the same mode shape $A(y)$, we have $A_{rms}^{th}=\left[\frac{1}{L}\int_0^L \frac{1-cos(2 \pi y/L)}{2}A_{max}^{th} dy\right]^{1/2}=\sqrt{\frac{3}{8}}A_{max}^{th}$.  Therefore, a sensitivity of $\frac{A_{max}^{th}}{\sqrt{f_m/Q}}\approx 35\ \textrm{fm}/\sqrt{\textrm{Hz}}$, about one order of magnitude smaller than our present $\sqrt{S_x}$, is needed to observe the thermal motion.  Further improvements on sensitivity can be made, e.g., by using a cold amplifier or decreasing the gap between the graphene and the gate.

Different driving forces usually result in different $Q$ values for mechanical resonators \cite{Eichler2011NatureNanotechnology, Singh2010Nanotechnology}. In order to single out factors affecting the $Q$ value, it is necessary to keep the driving force as fixed as possible. For this purpose, we carried out measurements where $V_{AC}$ was changed according to $V_{DC}$, keeping the product $V_{DC}\cdot V_{AV}$ constant; the prefactor in the drive $\frac{\partial C}{\partial D }|_{D=D_{eq}}$ changed only by 1\% according to our estimates. Fig. \ref{DuffingAndQ}(b) shows the resonance curves of sample \#2 under a constant driving force in the linear regime. The $Q$ values at different $V_{DC}$ obtained by Lorentzian fits are also plotted in Fig. \ref{DuffingAndQ}(b). The data shows a decreasing $Q(V_{DC})$ with increasing $|V_{DC}|$ while the amplitude $h$ of the response remain constant. As $Q\cdot h$ increases, this means that either the mode shape changes, or there is a slight increase in  $\frac{\partial C}{\partial D }|_{D=D_{eq}}$.

Typically, tension enhances the mechanical quality factor. In graphene stretched by electrical forces, displacement currents may induce considerable dissipation due to the large resistance of graphene.  The displacement current generated by $V_{DC}$ and time varying capacitance is given by $I_d(V_{DC})=\omega_m V_{DC} \delta C \cdot sin(\omega_m t)$. Since the resistance of gated graphene behaves as $R_G(V_{DC})=[g_0(1+\beta|V_{DC}|)]^{-1}$, the energy dissipated per period by the displacement current becomes $E_e(V_{DC})\approx \int_{0}^{2\pi/\omega_m}I_d^2 R_G dt=\frac{\pi \omega_m (\delta C)^2}{g_0}\cdot \frac{V_{DC}^2}{1+\beta|V_{DC}|}$; here $g_0$ is the minimum conductance and $\beta$ parametrizes the increase in conductance with $V_{DC}$.  In Fig. \ref{DuffingAndQ}(b), $\omega_m$ does not change much with $V_{DC}$, and the total energy stored in the resonator $E_t$ can be taken as a constant due to the equal heights of the resonance peaks. So with a $V_{DC}$-independent energy loss $E_m$ per period, the overall $Q$ value can be written as $Q(V_{DC})=\frac{2\pi E_t}{E_m+E_e(V_{DC})}$, or $$\frac{1}{Q(V_{DC})}=\frac{1}{Q_m}+\frac{\alpha V_{DC}^2}{1+\beta |V_{DC}|}, \eqno{(5)}$$ where $Q_m=Q(0)=2\pi E_t/E_m$ and $\alpha=\frac{\omega_m (\delta C)^2}{2E_t g_0}=\frac{(\frac{\partial C}{\partial D})^2|_{D=D_{eq}}}{m_{eff} \omega_m g_0}$ are constants.  With $g_0=23\ \mu\textrm{S}$ and $\beta=0.01$, Eq. (5) fits well with the experimental data, as shown by the red dashed curve in Fig. \ref{DuffingAndQ}(b).  For this sample, we obtain $m_{eff}=2.2 \times 10^{-17} \textrm{kg}$ by fitting $f_m(V_{DC})$ with Eq. (4).  For the effective mass density we obtain $\rho_{eff}=9.7 \rho_0$, where $\rho_0 =7.6 \times 10^{-19} \textrm{kg}/ \mu \textrm{m}^2$ is the ideal mass density of monolayer graphene.  Note that the suspended graphene conductance is rather insensitive to the gate voltage (i.e., $\beta$ is small) due to the absence of a dielectric layer between the gate and graphene. As a consequence, there is no need in Eq. (5) to take into account the Dirac point shift, which would bring only an insignificant asymmetry to the $Q(V_{DC})$ curve. The large $\rho_{eff}$ is probably a result of unpredictable adsorbates, while the small $g_0$ may arise due to bad contact resistance in addition to adsorbates.  Improvement on these numbers can be expected by annealing \cite{Bolotin2008SolidStateComm}. However, our method is easily generalized, for graphene mechanical resonators suspended on rigid target supports by using a similar non-flip assembly.

\begin{figure}
\includegraphics[width=\columnwidth]{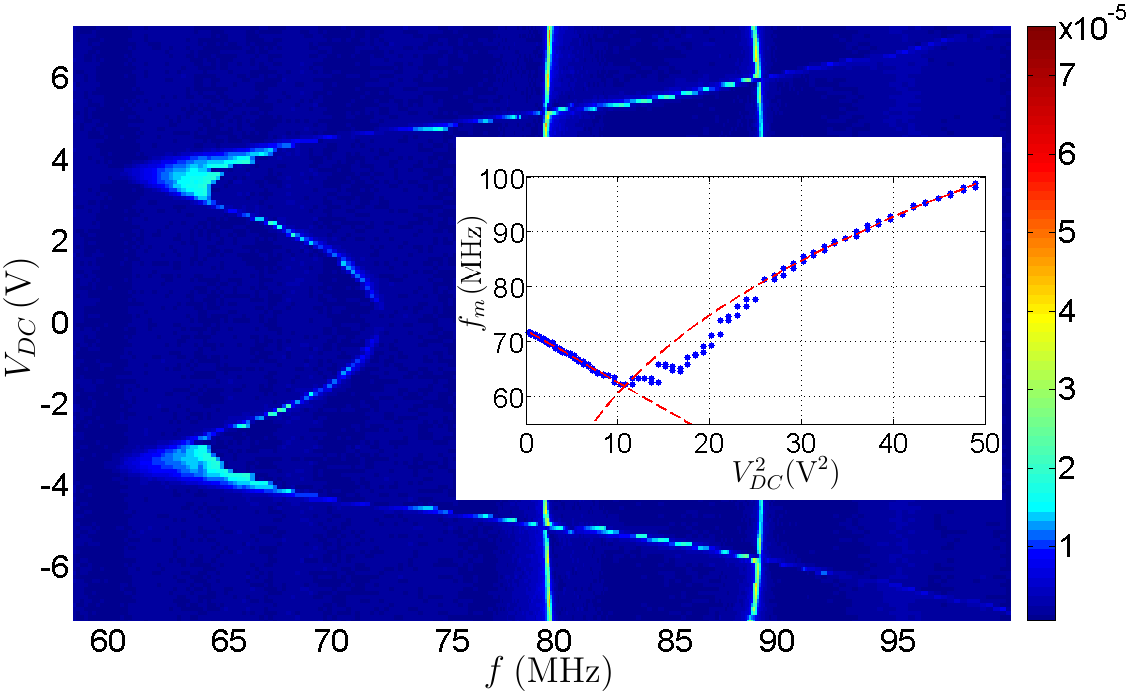}%
\caption{The mechanical resonance frequency $f_m$ as a function of DC gate voltage $V_{DC}$ for sample \#3.  The amplitude of the sideband $V_{\pm}$ is given by the color scale in Volts.  The non-monotonic, "W"-shaped curve shows a transition from the capacitor-dominated (low $|V_{DC}|$) regime to the tension-dominated (high $|V_{DC}|$) regime.  The other two straight traces with little $V_{DC}$ dependence are from the gold clamps with hanging ends like cantilevers.  Inset: the two red dashed lines are the two different fitting curves for low/high $|V_{DC}|$ regimes, respectively.}\label{W-shape}
\end{figure}

Complementary to the down-shifting behavior of Eq. (4), the deformation-induced tension in graphene leads to  up-shifting tendency at higher $|V_{DC}|$.
In Fig. \ref{W-shape}, we present data from a few-layer graphene sample \#3 with $L=W=1\ \mu$m, where the "W"-shape $f_m(V_{DC})$ curve shows a clear transition from the electric-force dominated regime to the tension dominated behavior.  In the high $|V_{DC}|$ regime, $f_m(V_{DC})$ fits the $f_m \propto V_{DC}^{2/3}$ model \cite{Chen2009NatureNanotechnology, Sapmaz2003PRB}, indicating a negligible intrinsic tension in the transferred graphene of this sample.  In the intermediate $|V_{DC}|$ regime, $f_m$ is less dependent on $V_{DC}$ and the resonance response curves become wider and asymmetric, implying some extra dissipation and nonlinear mechanics.  Analogous experimental signatures in this transition regime have been found in a recent work on nanowire resonators \cite{Solanki2010PRB}, where the authors attribute the widening of resonance to the mixing of different modes.  In our case of graphene, where no other modes were involved in the adjacent frequencies, a convincing theoretical model remains to be worked out.  In the low $|V_{DC}|$ regime, $f_m(V_{DC})$ fits well with the parabolic model of Eq. (4), which gives an effective mass $m_{eff} \sim 5 \times 10^{-19}$ kg for this sample, corresponding to $\rho_{eff}\sim 0.65 \rho_0$ when $ \frac{\partial^2 C}{\partial D^2 }|_{D=D_{eq}}$ is calculated from the parallel plate capacitance model. Although it is known that the effective mass of a clamped-clamped mechanical resonator can be smaller than the real mass of the suspended part \cite{LaHaye2004Science}, we attribute the small determined mass to edge effects that increase the actual value of $\frac{\partial^2 C}{\partial D^2 }$ than the calculated coefficient. In the case of clamped-clamped graphene, the unpredictable adsorbates and free edges make it more difficult to get an ideal effective mass due to the extremely small mass and thickness of graphene itself. Hence, we conclude that, adsorbates and edges must be carefully considered in both sample-making and modeling to determine the resonator mass correctly.

In conclusion, we have developed a micro-stamp technique that can be used to transfer and assemble suspended graphene into nanoscale mechanical resonators.  This new technique enabled us to take advantage of a localized gate in close proximity to the sample, which improved the sensitivity of the RF-cavity readout scheme by enhancing the capacitance ratio between the graphene device and the parasitic contributions.  We found Duffing effects of different sign on graphene samples, and measured the critical vibration amplitude at which the Duffing hysteresis emerges.  We observed "W"-shaped, nonmonotonic $f_m(V_{DC})$ curve on a few-layer graphene mechanical resonator, demonstrating a continuous transition from electrically dominated low $|V_{DC}|$ regime to tension-dominated high $|V_{DC}|$ regime.  In our experiments on a series of graphene samples, we obtained resonance frequencies up to 178 MHz, a sensitivity of about $0.3\ \textrm{pm}/\sqrt{\textrm{Hz}}$, and an effective resonator mass down to $\sim 10^{-18}$ kg.  The achieved combination of high frequency, high sensitivity, and low mass show that our micro-stamp technique and RF reflection measurement scheme hold promise for building novel, sensitive NEMS structures.  By using symmetrized electrical cavities\cite{Massel-ArXiv}, the sensitivity of our scheme can be further increased and the quantum limit is within reach with graphene mechanical resonators.

We thank T. Heikkil\"a, R. Khan, and D. Lyashenko for fruitful discussions. This work was supported by the Academy of Finland (contracts no. 132377 and 130058), the ERC contract FP7-240387, and by project RODIN FP7-246026. The work in the US was supported under NSF DMR-0908634.

\end{document}